%% file: conference_101719.tex
\documentclass[conference]{IEEEtran}
\IEEEoverridecommandlockouts
\usepackage{cite}
\usepackage{amsmath,amssymb,amsfonts}
\usepackage{algorithmic}
\usepackage{graphicx}
\usepackage{textcomp}
\usepackage{booktabs}
\usepackage{multirow}
\usepackage{xcolor}

\DeclareMathOperator*{\argmax}{arg\,max}
\usepackage{xcolor}
\def\BibTeX{{\rm B\kern-.05em{\sc i\kern-.025em b}\kern-.08em
    T\kern-.1667em\lower.7ex\hbox{E}\kern-.125emX}}
\begin{document}

\title{Improved Zero-Shot Audio Tagging \&  Classification
with Patchout Spectrogram Transformers}

\author{\IEEEauthorblockN{Paul Primus$^1$, Gerhard Widmer$^{1,2}$}
\IEEEauthorblockA{$^1$Institute of Computational Perception (CP-JKU)  \\
$^2$LIT Artificial Intelligence Lab\\
Johannes Kepler University, Austria}
}

\maketitle

\begin{abstract}
Standard machine learning models for tagging and classifying acoustic signals cannot handle classes that were not seen during training. Zero-Shot (ZS) learning overcomes this restriction by predicting classes based on adaptable class descriptions. This study sets out to investigate the effectiveness of self-attention-based audio embedding architectures for ZS learning. To this end, we compare the very recent patchout spectrogram transformer with two classic convolutional architectures. We evaluate these three architectures on three tasks and on three different benchmark datasets: general-purpose tagging on AudioSet, environmental sound classification on ESC-50, and instrument tagging on OpenMIC. Our results show that the self-attention-based embedding methods outperform both compared convolutional architectures in all of these settings. By designing training and test data accordingly, we observe that prediction performance suffers significantly when the `semantic distance' between training and new test classes is large, an effect that will deserve more detailed investigations.

\end{abstract}


\section{Introduction}

\begin{figure}[ht]
    \centering
    \def\svgwidth{0.75 \columnwidth}
    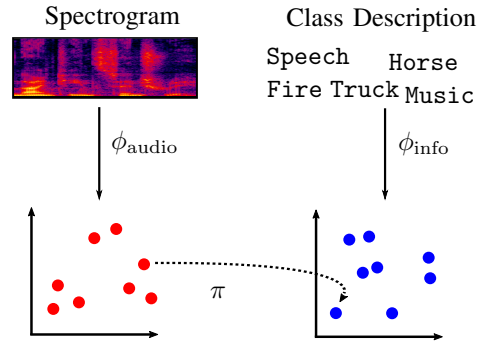
    \caption{Zero-Shot Learning for Audio in a nutshell: Audio and class description are transformed into separate embedding spaces via $\phi_\mathrm{audio}$ and $\phi_\mathrm{info}$, respectively. A non-linear cross-model projection $\pi$ maps audio embeddings close to their corresponding word embeddings. Classification and tagging are done based on distance metrics in the shared (blue) space.}
    \label{fig:zsl}
\end{figure}

Taggers and classifiers for audio signals are machine learning models that recognise certain acoustic events, such as the sounds of instruments, animals, or humans in sound recordings, on-line or off-line.
% \gw{"predict" is not the right word; "predict acoustic events" sounds like "predict that a dog will bark". }
Tagging problems are different from classifications problems in that they permit an arbitrary number of labels to be assigned to a single recording, whereas in classification tasks, precisely one label must be chosen. Current state-of-the-art methods for both classification and tagging are based on supervised learning of Convolutional Neural Networks (CNNs) or self-attention-based transformers. However, these classifiers are typically trained to distinguish between a fixed set of classes, rendering them useless for novel concepts that may appear later. \textit{Zero-Shot (ZS) learning} \cite{Larochelle2008} strives to overcome this issue by training generic classifiers that tag or classify based on distances in a joint data--label space. The basic idea is to transform input features and class characterizations into a common representation space such that the distance between related concepts is small. Descriptions or names of novel classes and audio queries can be projected into the same shared space, enabling classification and tagging via distance metrics, without ever having seen any labeled instances of this new class before. Figure \ref{fig:zsl} sketches the general idea. Universal classifiers like this have two compelling benefits over traditional classifiers: they require neither labeled training data for the novel class, nor any re-training.% \\

The main contribution of this study is to push the performance of ZS learning in the audio domain by capitalizing on state-of-the-art audio embedding models. We compare the time-honored VGGish embedding model used in previous works with a more recent CNN architecture and a self-attention-based audio spectrogram transformer to investigate if the supervised performance improvement also transfers to the ZS setting. Most studies on ZS learning for audio have focused on classification problems rather than tagging problems. We address this paucity by establishing the first baseline for ZS tagging on the \textit{AudioSet} \cite{Gemmeke2017} data set of acoustic recordings. To that end, we propose an evaluation scheme for ZS learning that considers the class imbalance in AudioSet. Additionally, we will evaluate the same general-purpose audio taggers, trained on AudioSet, in Zero-Shot classification and tagging on \textit{ESC-50} \cite{Piczak2015} and \textit{OpenMic} \cite{Humphrey2018}, respectively, without training on a single example from these data sets.

\section{Previous Work}

The first issue is how to map \textit{class information (labels)} into some embedding space. Previous work on ZS learning for audio has focused on two kinds of class descriptors: attribute vectors and textual information. Attribute vectors represent high-level classes as a combination of lower-level attributes. Choi et al. \cite{Choi2019}, for example, represent music genres as a combination of instruments. However, this requires specialized data sets labeled with attributes and may fail if a novel class cannot be adequately represented as a combination of those. More commonly used alternatives are general semantic representations that are learned from large text corpora via natural language models such as Word2Vec \cite{Pennington2014}, Glove \cite{Mikolov2013}, or BERT \cite{Devlin2018}, enabling us to potentially represent a large vocabulary or even full textual descriptions. However, these semantic spaces are not without limitations, as they sometimes fail to capture very specialized contexts, as we will see in our experiments. % \\
%%% \gw{GW: Bemerkung: Das waer was fuer spaetere genauere Untersuchungen, wenn das Paper fertig ist ..}

On the \textit{audio} input side, most existing zero-shot classifiers extract low dimensional representations from spectrograms via pre-trained audio embedding models. Pre-training is a common strategy, especially when labeled training data is scarce. CNN-based embedding models such as VGGish \cite{Hershey2017} and PANNs \cite{Kong2020} are the most widely adopted feature extractors. However, recent work \cite{Gong2021,Koutini2021}
used self-attention-based transformer architectures \cite{Dosovitskiy2020,Touvron2020} pre-trained on image and audio data sets and achieved state-of-the-art results on audio classification and tagging benchmark sets. %\\

There are only a handful of studies on ZS learning in the general audio domain; the majority of them has focused on single-label classification problems and used textual labels as class descriptors.
Islam et al. \cite{Islam2019} projected audio and textual class labels into separate lower-dimensional embedding spaces via a pre-trained Siamese Network \cite{BROMLEY1993} and Word2Vec, respectively, then mapped the audio embeddings into the semantic space 
% close to their corresponding label embedding
via a non-linear projection. New audio examples were classified as the closest candidate label in the semantic embedding space. Xie et al. \cite{Xie2019} used a pre-trained VGGish embedding model in a similar framework but with a linear cross-modal projection. Follow-up work \cite{Xie2021, Xie2021.2} mapped the audio embedding using factored linear and non-linear transformations, and also experimented with semantic embeddings from labels and textual descriptions generated with GloVe and BERT, respectively. Except for a single study in the music domain \cite{Choi2019}, previous studies have not dealt with audio tagging in much detail. Both 
% Islam et al. 
\cite{Islam2019} and
% Xie et al.
\cite{Xie2021, Xie2021.2} used parts of AudioSet \cite{Gemmeke2017}, a large data set designed for audio tagging, in their experiments. However, they discarded all recordings with multiple tags because they focused on single-label classification problems; by doing so, approximately 93\% of the AudioSet training data was not used. Moreover, all of these studies have used convolutional audio embedding models, such as the VGGish network; however, more promising approaches based on self-attention have recently been proposed; these are in the focus of the following investigation. %\\

\section{Spectrogram Embedding}

The audio embedding model $\phi_\mathrm{audio}:  \mathbb{R}^{f \times t} \rightarrow \mathbb{R}^m$ transforms the input spectrogram $x$ with $f$ frequency bins and $t$ time frames to an $m$-dimensional vector representation. In this work, we are going to compare three different architectures for the audio embedding: VGGish,  CNN14, and the Patchout faSt Spectrogram Transformer (PaSST)  \cite{Koutini2021}. % \\

\begin{table*}[ht]
\centering
\begin{tabular}{@{}c|cc|cccccc|c|c|c@{}}
\toprule
\multicolumn{1}{l|}{} & \multicolumn{1}{l}{\# params} & \multicolumn{1}{l|}{\# MEL} & fold 0 & fold 1 & fold 2 & \multicolumn{1}{l}{fold 3} & \multicolumn{1}{l}{fold 4} & \multicolumn{1}{l|}{avg} & \multicolumn{1}{l|}{ESC-50} & \multicolumn{1}{l|}{OpenMic} & \multicolumn{1}{l}{all} \\ \midrule
VGG                   & 75.0M                         & 64                          & 37.73  & 35.68  & 36.24  & 36.07                      & 37.69                      & 36.12                    &         37.71                    &           36.33                   &  38.31                       \\
CNN14                 & 77.5M                         & 128                         & 39.76  & 39.45  & 40.28  & 40.02                      & 41.96                      & 40.29                    &        40.11                     &      38.48                        &  40.86                       \\
PaSST                 & 86.1M                         & 128                         & 42.55  & 42.79  & 43.73  & 44.09                      & 45.11                      & 43.65                    & 43.49                       &      41.17                        & 45.23                   \\ \bottomrule
\end{tabular}
\caption{Overview of pre-trained embedding models. First section: model and input sizes. Remaining sections: results of supervised training in terms of mAP (\%) on the training classes for each experimental setup (fold 0-4, ESC-50, and OpenMIC) and the fully-supervised upper bound (all).}
\label{table:pretraining}
\end{table*}

The VGGish network \cite{Hershey2017} is named after the VGG architecture for image classification \cite{Simonyan2015} but is shallower than other VGG variants. It consists of 6 convolutional layers with batch normalization and ReLU activations and MaxPooling layers in between. Two fully connected layers follow the convolutional part with 4096 units and ReLUs activations. Due to the flattening operation before the fully-connected layers, the VGGish architecture has a fixed input size of $96$ time frames with $64$ MEL bins each, which means that longer inputs need to be split into chunks. % \\

CNN14 was used as the standard architecture in \cite{Kong2020} and consists of 12 convolutional layers and a single 2048-unit fully-connected layer. It further differs from VGGish in that it uses more kernels in the last layers and aggregates over the time and frequency dimension of the final feature maps, which permits inputs of arbitrary length. To harmonize the embedding sizes across all models, we modify the convolutional architectures so that they produce an output of size $768$. % \\

PaSST \cite{Koutini2021} is based on the vision transformer architecture first introduced for image classification and later re-used for audio tagging. Standard transformers operate on sequences of tokens and spectrogram transformers construct this sequence by dividing the 2D input into patches and converting each of them to a 1D vector. A learnable class token, which is later used for classification, is appended to the sequence, and the result is iteratively transformed via self-attention layers. Self-attention layers are unaware of the token's ordering; PaSST therefore adds a positional embedding to encode each token's time and frequency position. To reduce computational and memory requirements (which are both quadratic in sequence length), PaSST introduces \textit{patchout regularization}, which shortens the input sequence's length by stochastically dropping tokens, which in turn improves the performance compared to vanilla audio spectrogram transformers. In our experiments, we use \textit{structured patchout} \cite{Koutini2021} and non-overlapping spectrogram patches. The first part of Table \ref{table:pretraining} compares previously described architectures in terms of parameters and input size. % \\

\section{Semantic Embedding}
The semantic embedding $\phi_\mathrm{info}: \{0, 1\}^v \rightarrow \mathbb{R}^{n}$ maps a binary encoded label $c_i$ for class $i$ from a vocabulary of size $v$ to an $n$-dimensional embedding space in which related words are close to each other and directions capture variations in certain semantic dimensions. In this work, we consider two pre-trained models, namely Word2Vec \cite{Mikolov2013} and GloVe \cite{Pennington2014}. Word2Vec learns the semantic space by predicting a word's local context, i.e., the surrounding words in a given sentence. GloVe instead uses global word co-occurrence statistics of the entire text corpus to construct word vector representations. We choose Word2Vec and Glove models that are pre-trained on the Google News and the Common Crawl dataset, respectively. Both produce embedding vectors of size 300. To embed textual labels made out of multiple words (e.g., \textit{fire truck}), we first embed the words separately and then average the result.

\section{Cross-Modal Projection}
We use a non-linear cross-modal projection $\pi: \mathbb{R}^{m} \rightarrow \mathbb{R}^{n}$ to integrate the audio embeddings into the semantic space. $\pi$ is a fully-connected network with one hidden layer of 1024 units and GELU activations \cite{Hendrycks2016}. Inputs are normalized and dropout regularization \cite{Nitish2014} is applied on the intermediate representation to prevent overfitting. % \\

Putting all parts together, we then define the \textit{ZS audio tagger} that computes the class posterior probability for a class with descriptor $c$ and audio input $x$ as: 
$$P(c\mid x) = \hat{y} = \sigma \bigg ( \pi\big( \phi_\mathrm{audio}(x) \big)^T \phi_\mathrm{info}(c) \bigg) $$
where $\sigma: \mathbb{R} \rightarrow [0,1]$ is the sigmoid activation function that maps the logit output to a valid probability. For audio classification benchmarks, we predict the class with the highest class probability among the test classes $C_\mathrm{tst}$:

$$\hat{c} = \argmax_{c \in C_\mathrm{tst}} P(c \mid x) $$

The cross-modal projection is trained with the audio embedding model's pre-trained parameters frozen. We minimize the standard binary-cross entropy loss for all training classes $C_\mathrm{trn}$ and all instances in the training set $\mathcal{D}_\mathrm{trn}$.

\section{Experiments}

We train the ZS model variants on samples and classes included in AudioSet and evaluate them on three data sets: for general tagging on AudioSet, environmental sound classification on ESC-50, and tagging in the music domain on OpenMic. For each of these evaluation settings, the labels of the AudioSet classes that correspond to the classes in the test set are excluded during training. The following section details the data sets, the exact experimental setup, and the training.

\subsection{AudioSet Tagging Experiments}

First, we evaluate the methods' general ZS tagging performance on AudioSet \cite{Gemmeke2017}. This data set contains approximately 2 Million ten-second audio recordings labeled for 527 hierarchically organized classes and is one of the largest data sets for audio tagging available. The distribution of class tags is highly unbalanced: while \textit{Music} and \textit{Speech} are used more than a million times, narrower classes like \textit{Hi-Hat} or \textit{Babbling} appear only a few hundred or thousand times. We use the pre-defined data set split into a large, unbalanced set for training $\mathcal{D}_\mathrm{trn}$ and two smaller, more balanced sets for validation $\mathcal{D}_\mathrm{val}$ and testing $\mathcal{D}_\mathrm{tst}$. 
For the tagging experiments, we split Audioset's classes $C$ into disjoint subsets for training $C_\mathrm{trn}$ and testing $C_\mathrm{tst}$. More specifically, we split $C$ into five folds and perform cross-validation at that level to reduce the bias towards a specific selection of test classes. The number of tags in the folds is balanced as follows: We start with five empty folds, iterate over the classes in descending tag count order, and always assign the next class to the fold with the lowest number of overall tags. Classes \textit{Speech} and \textit{Music} are only used for training because they are tagged in more than $50 \%$ of the instances and thus hard to balance. Each resulting fold contains $105 \pm 1$ classes, approximately $374K$ tags, and approximately $315K$ labeled instances. We evaluate all combinations of previously introduced acoustic and semantic embedding models for general purpose ZS tagging on each class fold and $\mathcal{D}_\mathrm{tst}$ by train on instances in $\mathcal{D}_\mathrm{trn}$ using only the remaining classes.

\subsection{ESC-50 Classification Experiments}
We further evaluate the ZS classification capabilities on instances and classes in ESC-50 \cite{Piczak2015}, a data set for environmental sound classification. ESC-50 contains 20,000 five-second long audio recordings; the 50 classes (e.g., \textit{cat}) are pre-arranged into five parent categories (e.g., \textit{animals}) with ten classes each. In the results table, we will report the accuracy separately for each parent category and the dataset as a whole. For these experiments, we train on AudioSet's $\mathcal{D}_\mathrm{trn}$, after carefully excluding from AudioSet all classes that semantically overlap with class labels in ESC-50.
%We further evaluate the ZS classification capabilities on instances and classes in ESC-50 \cite{Piczak2015}, a data set for environmental sound classification. ESC-50 contains 20,000 five-second audio recordings, and each of them belongs to one out of 50 balanced classes \gw{(such as, e.g., ....)}. For these experiments, we train on AudioSet's $\mathcal{D}_\mathrm{trn}$, \gw{after} carefully excluding  \gw{from AudioSet} \gw{all classes that semantically overlap with class labels in ESC-50}.
% \pp{ja, das stimmt. es sind nicht alle classen 1-zu-1 in der audio set ontologie, aber ich hab manuell alle gleichen herausfgefiltert (z.B., 'air plane' und  'air craft')}}. 
%The data set is pre-arranged into five folds for evaluation; we, therefore, test on all instances. \gw{Den Satz versteh ich nicht. Was fuer eine Rolle spielen Folds und CrossVal hier?} \pp{ESC benchmarks werden normalerweise via 5-fold CV evaluiert. Weil wir aber eh nicht auf ESC trainieren, hab ich einfach auf dem ganzen set evaluiert.} 
%\gw{Dann wuerde ich gar nicht von "folds" sprechen, das verwirrt sehr, weil in der vorigen subsection von folds im kontext von crossval gesprochen wurde, und hier gibts ja eben keine CV. Wie waer'S mit "The dataset is pre-arranged into five subsets (called/relating to ... [hat das was mit "animal, natural.." in der Caption zu Table 2 zu tun (die ich auch nicht verstanden hab -- s.u.)?]). In the results table, we will report the .. both for the individual subsets, and for the dataset as a whole."}

\subsection{OpenMIC Tagging Experiments}
Finally, we assess the systems' performance on a more specific task, namely instrument tagging in the music domain. To this end, we use OpenMIC \cite{Humphrey2018}, a data set that contains 20,000 ten second long music recordings, partially labeled for the absence and presence of 20 instruments. 
%The number of positive and negative labels \gw{what are "positive" and "negative" labels? (Du hast es mir mal erklaert, aber der Leser weiss das nicht; ist es relevant? wenn ja, wie kann man das ganz kurz erklaeren?)} \pp{In OpenMIC sind nicht alle Instanzen für alle Instrumente annotiert. Dass die balanciert sind ist eher ein technisches Detail; hab ich hier nur erwähnt weil es für die random baseline interessant ist.} \gw{die unklarheit ist: was heisst "positive and negative labels"? als leser wuerde man erwarten, dass es nur labels gibt, die einem file zugeordnet oder nicht zugeordnet sind. aber "negative labels" is verwirrend.} in the data set is approximately balanced.
We again train the ZS models for these experiments on AudioSet's $\mathcal{D}_\mathrm{trn}$, carefully excluding either all instrument classes (condition $C \setminus C_\textrm{inst}$ in Table \ref{tab:mic}) or only those that appear as tags in OpenMIC (condition $C \setminus C_\textrm{MIC}$). For compatibility with other works using OpenMIC (e.g., \cite{Humphrey2018, Koutini2021}), we only use the $5,085$ instances in the test split of OpenMIC to evaluate the tagging performance. % \\

\subsection{Preprocessing \& Training}
%\gw{In der Section muss es noch Platzsparmoeglichkeiten geben; die ist laenger als die (interessanteren) vorigen subsections und etwas, was niemand im Detail lesen will (muss natuerlich da stehen, aber moeglichst kompakt).}

We convert the audio snippets into logarithmically weighted MEL-spectrograms with 128 bins (64 for VGGish), using an FFT window size of $800 \textrm{ms}$ and a hop size of $320 \mathrm{ms}$. PaSST parameters are initialized with weight from ImageNet \cite{Deng2010} models. To level the playing field, we do the same for VGGish and CNN14.
%To level the playing field \gw{was heisst das? ok, aber der leser weiss das nicht und kann deshalb den halbsatz nicht interpretieren.} \pp{PaSST ist mit den parametern von ViT initialisiert, deshalb hab ich das für die anderen Architekturen auch gemacht. ok, dann werd ich das noch ausformulieren} aber nicht zu lang und ausfuehrlich (platz!), ist ja kein wichtiger punkt., ok! we initialize the parameters of all three architectures by transferring the weights from pre-trained ImageNet models. 
We append a classification head to each architecture's output and pre-train the audio embedding models for 130 epochs on $\mathcal{D}_\mathrm{trn}$ to predict the classes in $C_\mathrm{trn}$. We use the AdamW \cite{Loshchilov2019} optimizer with an initial learning rate of $20^{-5}$, $\beta_1=0.9$, $\beta_2=0.99$, $\epsilon=10^{-8}$, and a weight decay of $10^{-4}$. The learning rate is exponentially increased to its initial value in the first five epochs and linearly decayed to $10^{-7}$ between epoch $50$ and $100$. We further use batch size $24$ and $32$ for PaSST and the convolutional models, respectively. To reduce the class imbalance during training, we use the balanced sampling strategy proposed in \cite{Kong2020}; this excludes samples that are not tagged for any of the classes in $C_\mathrm{train}$. We further use a range of augmentation techniques to prevent overfitting: Mix-Up regularization \cite{Zhang2018}, Specaugment \cite{Park2019}, time shifts, frequency shifts, and random gain augmentations. % \\

After pre-training, the classification head is replaced with the cross-modal projection, and training on classes $C_\mathrm{trn}$ in $\mathcal{D}_\mathrm{trn}$ is continued for ten more epochs with the embedding model's parameters frozen. We reset the learning rate to the initial value and use Audioset's $D_\mathrm{val}$ and approximately $10\%$ of the training classes for model selection. All reported ZS-experiments below are averaged over three runs. % \\

\section{Results}

Evaluation results of the pre-trained models on classes in $C_\mathrm{trn}$ and instances in $\mathcal{D}_{tst}$ are given in Table \ref{table:pretraining}. A comparison between the mean Average Precision (mAP) scores of the embedding models across all folds and additional training settings shows that VGGish is always worse than CNN14, which in turn is inferior to PaSST. This ranking is in line with the results reported in the original papers. However, we note a slight deviation from the metrics reported in the original works, which we attribute to different experimental setups. Furthermore, the table shows a consistent performance drop for the models trained on selected classes (fold 0-4, ESC-50, and OpenMIC) compared to those which are trained with all class tags (all). This is expected because multiple classes are dropped for training, and the available training data is reduced.

\subsection{AudioSet Tagging Results}
The results of the tagging experiments on AudioSet are given in Table \ref{tabel:audioset}. Although the mAP drops in the ZS setting by more than 30 pp. compared to the supervised upper bound in Table \ref{table:pretraining}, all outcomes are still significantly above the random baseline's mAP score ($0.4\%$). PaSST consistently outperforms both VGGish and PANN across all folds and semantic embedding methods, and we observe a similar performance gap between VGGish and CNN14. Moreover, Word2Vec and Glove perform roughly on par, regardless of the audio embedding method; however, Glove seems to have the edge on average. The test classes' AP improvement over the random baseline correlates weakly with proximity to their nearest neighbor in the semantic space ($r=0.22$), which could indicate that training with classes that are semantically close to the test classes is advantageous.

\begin{table}
\begin{tabular}{@{}lccclll@{}}
\toprule
                            & \multicolumn{3}{c}{Word2Vec}               & \multicolumn{3}{c}{GloVe} \\ \midrule
                            & VGGish & CNN14 & PaSST                     & VGGish  & CNN14  & PaSST  \\ \midrule
\multicolumn{1}{l|}{fold 0} & 8.91   & 9.82  & \multicolumn{1}{c|}{\textbf{11.12}} & 9.01    &  9.92      &  \textbf{10.96}      \\
\multicolumn{1}{l|}{fold 1} & 7.76   & 8.72  & \multicolumn{1}{c|}{\textbf{9.60}} &   7.71      &  8.85      &  \textbf{9.85} \\
\multicolumn{1}{l|}{fold 2} & 9.15  & 10.41  & \multicolumn{1}{c|}{\textbf{10.44}} &  9.15       &   10.27     &    \textbf{10.62}    \\
\multicolumn{1}{l|}{fold 3} & 6.99  &  7.88 & \multicolumn{1}{c|}{\textbf{8.95}} &  7.01       &  7.88     &    \textbf{8.89}    \\
\multicolumn{1}{l|}{fold 4} & 6.67 & 7.67  & \multicolumn{1}{c|}{\textbf{8.29}} &   6.75      &   7.73     &  \textbf{8.27}      \\ \midrule
mean                        &  7.90      &   8.90    & \textbf{9.68}                      &  7.93       &    8.93 &  \textbf{9.72}      \\ \bottomrule
\end{tabular}

\caption{mAP (\%) on the zero-shot test classes for each of the five folds and both semantic embedding methods Word2Vec and GloVe.}
\label{tabel:audioset}
\end{table}

\subsection{ESC-50 Classification Results}
We now move on to discuss the results of the classification experiments, which are summarized in Table \ref{tab:esc}. We evaluate the classification performance using Word2Vec word embedding category-wise and among all classes as ten-way and fifty-way classification problems, respectively. All results are significantly above the random baseline's accuracy score. A direct comparison between the embedding model reveals the same trend as in the previous experiments: PaSST outperforms both PANN in the fifty-way setting and for all categories except the Animal group. We observe a similar performance gap between VGGish and PaSST.
We further compare the results of our VGGish based model to the values reported in \cite{Xie2019}. Although the embedding models share roughly the same architecture and are tested on the same data set, the experimental setup is quite unlike. To highlight some differences: we pre-train custom VGGish embedding model and carefully exclude the test classes labels; the size of the embedding is different; we train the cross-modal projection exclusively on AudioSet and transfer our model's initial parameters from ImageNet models. In a direct comparison, we observe a slight deterioration ($< 1$ pp.) for categories Natural and Human, and significant improvements ($> 16$ pp.) for categories Animal, Domestic, and Exterior. This large discrepancy could be attributed to the higher number of classes used to train the cross-modal embedding.

\begin{table}[]
\centering
\begin{tabular}{@{}lcccc@{}}
\toprule
         & VGGish \cite{Xie2019}       & VGGish               & CNN14                & PaSST                                     \\ \midrule
Animal   & 22.2                 &            38.5            &    \textbf{45.8}                  & 43.8                                            \\
Natural  & 39.7                 &           38.5           &   42.2                   & \textbf{48.3}                                            \\
Human    & 22.0                 &         21.5             &     30.4                 & \textbf{35.0}                                            \\
Domestic & 17.0                 &         53.8             &     54.3                 & \textbf{59.0}                                           \\
Exterior & 29.2                 &         50.8            &         56.7             & \textbf{59.0}                                           \\ \midrule
All      & -                    &      31.5                &        35.8              & \textbf{40.5}                                           \\ \bottomrule
\end{tabular}
\caption{Top-1 Classification Accuracy (\%) for ESC-50, evaluated for each of the five parent categories (10-way classification.}
\label{tab:esc}
\end{table}

\subsection{OpenMIC Tagging Results}
Table \ref{tab:mic} details the evaluation results for the instrument tagging experiments on OpenMIC using Word2Vec word embedding. All results are above the random baseline's mAP score of $40\%$. When the ZS model is trained without any instrument class labels ($C \setminus C_\mathrm{inst}$), VGGish and PaSST perform on par, and, surprisingly, CNN14's performs worse than both. However, if the ZS model is trained using some instrument class labels ($C \setminus C_\mathrm{MIC}$), the tagging performance increases for all embedding models ($> 7$ pp.), and we observe the usual ranking. A comparison to the ZS tagger trained on all 527 classes in AudioSet ($C$, including the 20 instrument classes in OpenMIC) suggests that there is still improvement potential ($>10$ pp.).

\begin{table}[]
\centering
\begin{tabular}{@{}llll@{}}
\toprule
   $C_{\mathrm{train}}$              & VGGish   & CNN14   & PaSST  \\ \midrule
$C \setminus C_\textrm{inst}$   & 60.32    & 58.77   & \textbf{60.38}  \\
$C \setminus C_\textrm{MIC}$ &    67.43      &   68.35      &  \textbf{69.04}      \\
$C$  & 77.31    & 78.7   & \textbf{80.47}  \\ \bottomrule
\end{tabular}
\caption{mAP (\%) for instrument tagging on OpenMIC, with no, selected, or all instrument class labels used for training.}
\label{tab:mic}
\end{table}

\section{Conclusion}
This study set out to analyze whether the performance gains of self-attention-based embedding models for supervised audio tagging and classification also transfer to a ZS settings. To this end, we compared the performance of convolutional- and transformer-based audio embedding models on AudioSet, ESC-50, and OpenMIC. The results of our experiments favor the self-attention model over both compared convolutional architectures; however, this doesn't necessarily imply that self-attention architectures have an intrinsic advantage for the task at hand, and more recent convolutional architectures such as ConvNeXt \cite{Zhuang2022} could yield even better results.
We tested both Word2Vec and Glove for class label embedding and observed no significant difference between them. Furthermore, we evaluated the ZS taggers in a specialized context (testing on new classes (\textit{musical instruments}) that are very different, semantically, from the classes in the training data (\textit{animals} and \textit{engines})) and observed a significant performance drop in these cases; the issue of semantic distance between concepts definitely deserves further and deeper study.
Generally, the considerable performance gap between the seen and unseen classes in all settings suggests that there is still plenty of room for future work.
\bibliographystyle{ieeetr}
\bibliography{conference_101719}

\end{document}

%% file: image.pdf_tex
%% Creator: Inkscape 1.0.2 (1.0.2+r75+1), www.inkscape.org
%% Accompanies image file 'image.pdf' (pdf, eps, ps)
\begingroup%
  \makeatletter%
  \providecommand\color[2][]{%
    \errmessage{(Inkscape) Color is used for the text in Inkscape, but the package 'color.sty' is not loaded}%
    \renewcommand\color[2][]{}%
  }%
  \providecommand\transparent[1]{%
    \errmessage{(Inkscape) Transparency is used (non-zero) for the text in Inkscape, but the package 'transparent.sty' is not loaded}%
    \renewcommand\transparent[1]{}%
  }%
  \providecommand\rotatebox[2]{#2}%
  \newcommand*\fsize{\dimexpr\f@size pt\relax}%
  \newcommand*\lineheight[1]{\fontsize{\fsize}{#1\fsize}\selectfont}%
  \ifx\svgwidth\undefined%
    \setlength{\unitlength}{569.60086765bp}%
    \ifx\svgscale\undefined%
      \relax%
    \else%
      \setlength{\unitlength}{\unitlength * \real{\svgscale}}%
    \fi%
  \else%
    \setlength{\unitlength}{\svgwidth}%
  \fi%
  \global\let\svgwidth\undefined%
  \global\let\svgscale\undefined%
  \makeatother%
  \begin{picture}(1,0.67219146)%
    \lineheight{1}%
    \setlength\tabcolsep{0pt}%
    \put(0,0){\includegraphics[width=\unitlength,page=1]{image.pdf}}%
    \put(0.78691043,0.47404759){\color[rgb]{0,0,0}\makebox(0,0)[lt]{\lineheight{1.25}\smash{\begin{tabular}[t]{l}$\mathtt{Music}$\end{tabular}}}}%
    \put(0.50913778,0.55246914){\color[rgb]{0,0,0}\makebox(0,0)[lt]{\lineheight{1.25}\smash{\begin{tabular}[t]{l}$\mathtt{Speech}$\end{tabular}}}}%
    \put(0.51038776,0.48750893){\color[rgb]{0,0,0}\makebox(0,0)[lt]{\lineheight{1.25}\smash{\begin{tabular}[t]{l}$\mathtt{Fire\;Truck}$\end{tabular}}}}%
    \put(0,0){\includegraphics[width=\unitlength,page=2]{image.pdf}}%
    \put(0.39622363,0.08322445){\color[rgb]{0,0,0}\makebox(0,0)[lt]{\lineheight{1.25}\smash{\begin{tabular}[t]{l}$\pi$\end{tabular}}}}%
    \put(0.19834214,0.38252045){\color[rgb]{0,0,0}\makebox(0,0)[lt]{\lineheight{1.25}\smash{\begin{tabular}[t]{l}$\phi_\mathrm{audio}$\end{tabular}}}}%
    \put(0.7724282,0.38139204){\color[rgb]{0,0,0}\makebox(0,0)[lt]{\lineheight{1.25}\smash{\begin{tabular}[t]{l}$\phi_\mathrm{info}$\end{tabular}}}}%
    \put(4.26098802,1.82396843){\color[rgb]{0,0,0}\makebox(0,0)[lt]{\begin{minipage}{2.63046094\unitlength}\raggedright \end{minipage}}}%
    \put(0,0){\includegraphics[width=\unitlength,page=3]{image.pdf}}%
    \put(0.06567421,0.63310172){\color[rgb]{0,0,0}\makebox(0,0)[lt]{\lineheight{1.25}\smash{\begin{tabular}[t]{l}Spectrogram\end{tabular}}}}%
    \put(0.54902777,0.62911143){\color[rgb]{0,0,0}\makebox(0,0)[lt]{\lineheight{1.25}\smash{\begin{tabular}[t]{l}Class Description\end{tabular}}}}%
    \put(0.75388666,0.54132612){\color[rgb]{0,0,0}\makebox(0,0)[lt]{\lineheight{1.25}\smash{\begin{tabular}[t]{l}$\mathtt{Horse}$\end{tabular}}}}%
    \put(0,0){\includegraphics[width=\unitlength,page=4]{image.pdf}}%
  \end{picture}%
\endgroup%

%% file: conference_101719.bbl
\begin{thebibliography}{10}

\bibitem{Larochelle2008}
H.~Larochelle, D.~Erhan, and Y.~Bengio, ``Zero-data learning of new tasks,'' in
  {\em Proc. of the 23rd {AAAI} Conf. on Artificial Intelligence, {AAAI} 2008,
  Chicago, IL}.

\bibitem{Gemmeke2017}
J.~F. Gemmeke, D.~P.~W. Ellis, D.~Freedman, A.~Jansen, W.~Lawrence, R.~C.
  Moore, M.~Plakal, and M.~Ritter, ``Audio {S}et: An ontology and human-labeled
  dataset for audio events,'' in {\em 2017 {IEEE} Int. Conf. on Acoustics,
  Speech and Signal Processing, {ICASSP} 2017, New Orleans, LA}, 2017.

\bibitem{Piczak2015}
K.~J. Piczak, ``{ESC:} dataset for environmental sound classification,'' in
  {\em Proc. of the 23rd Annual {ACM} Conf. on Multimedia Conf., {MM} 2015,
  Brisbane, Australia}.

\bibitem{Humphrey2018}
E.~Humphrey, S.~Durand, and B.~McFee, ``Openmic-2018: An open data-set for
  multiple instrument recognition,'' in {\em Proc. of the 19th Int. Society for
  Music Information Retrieval Conf., {ISMIR} 2018, Paris, France}.

\bibitem{Choi2019}
J.~Choi, J.~Lee, J.~Park, and J.~Nam, ``Zero-shot learning for audio-based
  music classification and tagging,'' in {\em Proc. of the 20th Int. Society
  for Music Information Retrieval Conf., {ISMIR} 2019, Delft, Netherlands}.

\bibitem{Pennington2014}
J.~Pennington, R.~Socher, and C.~D. Manning, ``Glove: Global vectors for word
  representation,'' in {\em Proc. of the 2014 Conf. on Empirical Methods in
  Natural Language Processing, {EMNLP} 2014, Doha, Qatar}.

\bibitem{Mikolov2013}
T.~Mikolov, K.~Chen, G.~Corrado, and J.~Dean, ``Efficient estimation of word
  representations in vector space,'' in {\em 1st Int. Conf. on Learning
  Representations, {ICLR} 2013, Workshop Track Proc., Scottsdale, AZ}.

\bibitem{Devlin2018}
J.~Devlin, M.~Chang, K.~Lee, and K.~Toutanova, ``{BERT:} pre-training of deep
  bidirectional transformers for language understanding,'' in {\em Proc. of the
  2019 Conf. of the North American Chapter of the Association for Computational
  Linguistics: Human Language Technologies, {NAACL-HLT} 2019, Volume 1 (Long
  and Short Papers), Minneapolis, MN}.

\bibitem{Hershey2017}
S.~Hershey, S.~Chaudhuri, D.~P.~W. Ellis, J.~F. Gemmeke, A.~Jansen, R.~C.
  Moore, M.~Plakal, D.~Platt, R.~A. Saurous, B.~Seybold, M.~Slaney, R.~J.
  Weiss, and K.~W. Wilson, ``{CNN} architectures for large-scale audio
  classification,'' in {\em 2017 {IEEE} Int. Conf. on Acoustics, Speech and
  Signal Processing, {ICASSP} 2017, New Orleans, LA}.

\bibitem{Kong2020}
Q.~Kong, Y.~Cao, T.~Iqbal, Y.~Wang, W.~Wang, and M.~D. Plumbley, ``{PANNs}:
  Large-scale pretrained audio neural networks for audio pattern recognition,''
  {\em {IEEE} {ACM} Trans. Audio Speech Lang. Process.}, vol.~28, 2020.

\bibitem{Gong2021}
Y.~Gong, Y.~Chung, and J.~R. Glass, ``{AST:} audio spectrogram transformer,''
  {\em CoRR}, vol.~abs/2104.01778, 2021.

\bibitem{Koutini2021}
K.~Koutini, J.~Schl{\"{u}}ter, H.~Eghbal{-}zadeh, and G.~Widmer, ``Efficient
  training of audio transformers with patchout,'' {\em CoRR},
  vol.~abs/2110.05069, 2021.

\bibitem{Dosovitskiy2020}
A.~Dosovitskiy, L.~Beyer, A.~Kolesnikov, D.~Weissenborn, X.~Zhai,
  T.~Unterthiner, M.~Dehghani, M.~Minderer, G.~Heigold, S.~Gelly, J.~Uszkoreit,
  and N.~Houlsby, ``An image is worth 16x16 words: Transformers for image
  recognition at scale,'' in {\em 9th Int. Conf. on Learning Representations,
  {ICLR} 2021, Virtual Event, Austria}.

\bibitem{Touvron2020}
H.~Touvron, M.~Cord, M.~Douze, F.~Massa, A.~Sablayrolles, and H.~J{\'{e}}gou,
  ``Training data-efficient image transformers {\&} distillation through
  attention,'' in {\em Proc. of the 38th Int. Conf. on Machine Learning, {ICML}
  2021, Virtual Event}, vol.~139.

\bibitem{Islam2019}
M.~T. Islam and S.~Nirjon, ``Soundsemantics: exploiting semantic knowledge in
  text for embedded acoustic event classification,'' in {\em Proc. of the 18th
  Int. Conf. on Information Processing in Sensor Networks, {IPSN} 2019,
  Montreal, Canada}.

\bibitem{BROMLEY1993}
J.~Bromley, I.~Guyon, Y.~LeCun, E.~S{\"{a}}ckinger, and R.~Shah, ``Signature
  verification using a siamese time delay neural network,'' in {\em Advances in
  Neural Information Processing Systems 6, {NIPS}, Denver, CO}, 1993.

\bibitem{Xie2019}
H.~Xie and T.~Virtanen, ``Zero-shot audio classification based on class label
  embeddings,'' in {\em 2019 {IEEE} Workshop on Applications of Signal
  Processing to Audio and Acoustics, {WASPAA} 2019, New Paltz, NY}.

\bibitem{Xie2021}
H.~Xie and T.~Virtanen, ``Zero-shot audio classification via semantic
  embeddings,'' {\em {IEEE} {ACM} Trans. Audio Speech Lang. Process.}, vol.~29,
  2021.

\bibitem{Xie2021.2}
H.~Xie, O.~R{\"{a}}s{\"{a}}nen, and T.~Virtanen, ``Zero-shot audio
  classification with factored linear and nonlinear acoustic-semantic
  projections,'' in {\em {IEEE} Int. Conf. on Acoustics, Speech and Signal
  Processing, {ICASSP} 2021, Toronto, Canada}.

\bibitem{Simonyan2015}
K.~Simonyan and A.~Zisserman, ``Very deep convolutional networks for
  large-scale image recognition,'' in {\em 3rd Int. Conf. on Learning
  Representations, {ICLR} 2015, Conf. Track Proc., San Diego, CA}.

\bibitem{Hendrycks2016}
D.~Hendrycks and K.~Gimpel, ``Bridging nonlinearities and stochastic
  regularizers with gaussian error linear units,'' {\em CoRR},
  vol.~abs/1606.08415, 2016.

\bibitem{Nitish2014}
N.~Srivastava, G.~E. Hinton, A.~Krizhevsky, I.~Sutskever, and R.~Salakhutdinov,
  ``Dropout: a simple way to prevent neural networks from overfitting,'' {\em
  J. Mach. Learn. Res.}, vol.~15, no.~1, 2014.

\bibitem{Deng2010}
J.~Deng, W.~Dong, R.~Socher, L.~Li, K.~Li, and L.~Fei{-}Fei, ``{ImageNet}: {A}
  large-scale hierarchical image database,'' in {\em 2009 {IEEE} Computer
  Society Conf. on Computer Vision and Pattern Recognition {(CVPR} 2009),
  Miami, FL}.

\bibitem{Loshchilov2019}
I.~Loshchilov and F.~Hutter, ``Decoupled weight decay regularization,'' in {\em
  7th Int. Conf. on Learning Representations, {ICLR} 2019, New Orleans, LA}.

\bibitem{Zhang2018}
H.~Zhang, M.~Ciss{\'{e}}, Y.~N. Dauphin, and D.~Lopez{-}Paz, ``Mixup: Beyond
  empirical risk minimization,'' in {\em 6th Int. Conf. on Learning
  Representations, {ICLR} 2018, Conf. Track Proc., Vancouver, Canada}.

\bibitem{Park2019}
D.~S. Park, W.~Chan, Y.~Zhang, C.~Chiu, B.~Zoph, E.~D. Cubuk, and Q.~V. Le,
  ``{SpecAugment}: {A} simple data augmentation method for automatic speech
  recognition,'' in {\em 20th Annual Conf. of the Int. Speech Communication
  Association, Interspeech 2019, Graz, Austria}.

\bibitem{Zhuang2022}
Z.~Liu, H.~Mao, C.~Wu, C.~Feichtenhofer, T.~Darrell, and S.~Xie, ``A convnet
  for the 2020s,'' {\em CoRR}, vol.~abs/2201.03545, 2022.

\end{thebibliography}
